\documentclass{svjour2}                    
\smartqed  
\usepackage{graphicx}
\begin{document}

\title{Galactic rotation curves inspired by a
noncommutative-geometry background}
\titlerunning{Galactic rotation curves}

\author{F. Rahaman \and Peter K. F.
Kuhfittig \and K. Chakraborty \and A. A. Usmani \and Saibal Ray}
\authorrunning{Rahaman \and Kuhfittig \and Chakraborty \and Usmani \and Ray}

\institute{F. Rahaman \at Department of Mathematics, Jadavpur
University, Kolkata 700 032, West Bengal, India\\
\email{farook\_rahaman@yahoo.com} \and Peter K. F. Kuhfittig \at
Department of Mathematics, Milwaukee School of Engineering,
Milwaukee, Wisconsin 53202-3109, USA\\ \email{kuhfitti@msoe.edu}
\and K. Chakraborty \at Department of Physics, Government Training
College, Hooghly - 712103, West Bengal, India\\
\email{kchakraborty28@yahoo.com} \and A. A. Usmani \at Department
of Physics, Aligarh Muslim University, Aligarh 202 002, Uttar
Pradesh, India.\\ \email{anisul@iucaa.ernet.in} \and Saibal Ray
\at Department of Physics, Government College of Engineering and
Ceramic Technology, Kolkata 700 010, West Bengal, India \\
\email{saibal@iucaa.ernet.in}}

\date{Received: date / Accepted: date}

\maketitle

\begin{abstract}
This paper discusses the observed flat rotation curves of galaxies
in the context of noncommutative geometry.  The energy density of
such a geometry is diffused throughout a region due to the
uncertainty encoded in the coordinate commutator.  This intrinsic
property appears to be sufficient for producing stable circular
orbits, as well as attractive gravity, without the need for dark
matter.


\end{abstract}

\section{Introduction}
The inability to account for stellar motions in the outer regions
of galaxies has led to the hypothesis that galaxies and even
clusters of galaxies are pervaded by dark matter
\cite{Oort32,fZ33,fZ37}. This hypothesis has been confirmed by
observing the flatness of galactic rotation curves
\cite{kF70,RR73,OPY74,EKS74,RTF78,SR01}. The composition of such
matter, if it exists, has remained unknown.

The dark-matter problem originated in the measurement of the
tangential velocity $v^{\phi}$ of stable circular orbits of
hydrogen clouds in the outer regions of the halo. To explain the
observed constant velocity, it is assumed that the decrease in the
energy density is proportional to $r^{-2}$, where $r$ is the
distance from the center of the galaxy.

A number of candidates for dark matter have been proposed. The
most favored is the cold dark matter (CDM) paradigm
\cite{ESM90,aP04}. Another strong possibility is the $\Lambda$-CDM
model that is related to the accelerated expansion of the Universe
\cite{mT04a,mT04b}. For a summary of some alternative theories,
such as scalar-tensor and brane-world models, see Rahaman, et al.
\cite{fR07,fR08,fR10}.

In this paper we study the dark-matter problem from a completely
different perspective: an important outcome of string theory is
the realization that coordinates may become noncommuting operators
on a $D$-brane \cite{eW96,SW99}. The result is a discretization of
spacetime due to the commutator $[\bf{x}^{\mu},
\bf{x}^{\nu}]$$=i\,\theta^{\mu\nu}$, where $\theta^{\mu\nu}$ is an
antisymmetric matrix. Noncommutativity replaces point-like
structures by smeared objects \cite{SS03}, suggesting the
possibility of eliminating the divergences that normally appear in
general relativity.  The smearing effect is accomplished by using
a Gaussian distribution of minimal length $\sqrt{\alpha}$ instead
of the Dirac-delta function. More precisely, the energy density of
the static and spherically symmetric smeared and particle-like
gravitational source has the following form \cite{NSS06}:
\begin{equation}\label{E:density1}
\rho^* = \frac{M}{(4 \pi \alpha)^{\frac{3}{2}}}~exp
\left(-\frac{r^2}{4\alpha}\right),
\end{equation}
where the mass $M$ is diffused throughout a region of linear
dimension $\sqrt{\alpha}$ due to the uncertainty.  The
noncommutativity is an intrinsic geometric property of the
manifold, and not of its matter content, and can be taken into
account by keeping the standard form of the Einstein tensor on
the left-hand side of the field equations.  The right-hand side
is modified, however, by introducing a new energy-momentum
tensor as a gravitational source.

Taking the flat rotation curves as input, this model predicts
both stable circular orbits and attractive gravity in a
typical galaxy.

\section{The model}
In this paper the metric for a static spherically symmetric
spacetime  is taken as
\begin{equation}\label{E:line1}
ds^{2}=-e^{\nu(r)}dt^{2}+e^{\lambda(r)}dr^{2}
   +r^{2}(d\theta^{2}+ sin^{2}\theta \,d\phi^{2}),
\end{equation}
where the functions of the radial coordinate $r$,$\ \nu(r)$ and
$\lambda(r)$, are the metric potentials.

Now we consider the model with a maximally localized source
of energy. Here the Einstein equations can be written as
\begin{equation}
G_{\mu\nu}=   8 \pi G  T_{\mu\nu}.
\end{equation}
(We assume that $c=1$). The most general energy momentum tensor
compatible with static spherical symmetry is
\begin{equation}
T_\nu^\mu=diag( -\rho,~ p_r,~ p_t,~ p_t).
\end{equation}
For the metric (\ref{E:line1}), the Einstein field equations are
\begin{equation}\label{E:Einstein1}
e^{-\lambda}\left[\frac{\lambda^{\prime}}{r}-\frac{1}{r^{2}}\right]
+\frac{1}{r^{2}} = 8\pi G \rho,
\end{equation}
\begin{equation}\label{E:Einstein2}
e^{-\lambda}\left[\frac{1}{r^{2}}+\frac{\nu^{\prime}}{r}\right]
-\frac {1}{r^{2}} = 8\pi G p_r,
\end{equation}
 \begin{equation}\label{E:Einstein3}
\frac{1}{2}e^{-\lambda}\left[\frac{1}{2}(\nu^{\prime})^{2}+\nu^{\prime
\prime}-\frac{1}{2}\lambda^{\prime}\nu^{\prime}+\frac{1}{r}({\nu^{\prime
}-\lambda^{\prime}})\right] = 8\pi G p_t.
\end{equation}

\section{The solutions}\label{S:solutions}
Using the observed flat rotation curves as a starting point, it is
well known \cite{fR08,NVM09} that this condition gives the
solution
\begin{equation}\label{E:flat}
e^{\nu}= B_0 r^l,
\end{equation}
where $l$ is given by $l=2v^{2\phi}$ and $B_0$ is an integration
constant.  (For a derivation, see Ref. \cite{fR10}.) According to
Matos, Guzman and Lopez \cite{MGL00}, the observed rotation curve
profile in the presumed dark matter dominated region is such that
the rotational velocity $v^{\phi}$ becomes approximately constant
with $v^{\phi} \sim 300 $ km/s ($ \sim 10^{-3}$) for a typical
galaxy.  So $l=0.000001$, as shown by Nandi et al. \cite{kN09},
and we likewise assume large distances measured in kpc from the
galactic center.

Using equation (\ref{E:density1}), equation (\ref{E:Einstein1})
yields
\begin{equation}\label{E:lambda}
e^{-\lambda}=  1 - \frac{2 m^*(r) }{r},
\end{equation}
where

\begin{equation}\label{E:mass1}
m^*(r) = \frac{2M}{\sqrt{\pi}}~\gamma \left(\frac{3}{2} ,
\frac{r^2}{4\alpha}\right)
   =\frac{2M}{\sqrt{\pi}}\int_0^{r^2/4\alpha}
      \sqrt{t}~e^{-t}dt
\end{equation}
and
\begin{equation}\label{E:gamma1}
\gamma \left( \frac{3}{2}, \frac{r^2}{4\alpha}\right)  =
\int_0^{r^2/4\alpha}\sqrt{t}~e^{-t}dt
\end{equation}
is the lower incomplete gamma function \cite{NSS06}.  The
classical Schwarzschild mass is recovered in the limit
as $r/\sqrt{\alpha}\rightarrow\infty$.  Furthermore, new
physics can only be expected if $r\approx \sqrt{\alpha}$.

The mass $M$ could be a diffused centralized object such
as a wormhole \cite{LG08} or a gravastar \cite{LG10}.
Since we are interested in rotation curves at some fixed
distance $r=R_0$ from the center, we  need to consider
instead a spherical shell of radius $r=R_0$.  So instead
of a smeared particle-like object, we get a smeared shell.
We will therefore need to translate the above curves as
follows:
\begin{equation}\label{E:density2}
\rho = \frac{M}{(4 \pi \alpha)^{\frac{3}{2}}}~exp
\left(-\frac{(r-R_0)^2}{4\alpha}\right)
\end{equation}
and
\begin{equation}\label{E:mass2}
m(r) = \frac{2M}{\sqrt{\pi}}~\gamma \left( \frac{3}{2} ,
\frac{(r-R_0)^2}{4\alpha}\right).
\end{equation}
These replace $\rho^*$ and $m^*(r)$. As before
\begin{equation}\label{E:gamma2}
\gamma \left( \frac{3}{2}, \frac{(r-R_0)^2}{4\alpha}\right)
=\int_0^{(r-R_0)^2/4\alpha}\sqrt{t}~e^{-t}dt
\end{equation}
\noindent is the lower incomplete gamma function.  It is important
to note that Eq. (\ref{E:gamma2}) consists of a pure translation
of Eq. (\ref{E:gamma1}) and therefore remains an appropriate model
(see Fig. 1). Here $M$ denotes the mass of a spherical shell of
radius $r=R_0$. Once again, the classical Schwarzschild limit is
recovered as $(r-R_0)/\sqrt{\alpha}\rightarrow\infty$, while new
physics can only be expected if $r-R_0\approx\sqrt{\alpha}$.

\begin{figure}[tbp]
\begin{center}
\vspace{0.5cm}
\includegraphics[width=0.5\textwidth]{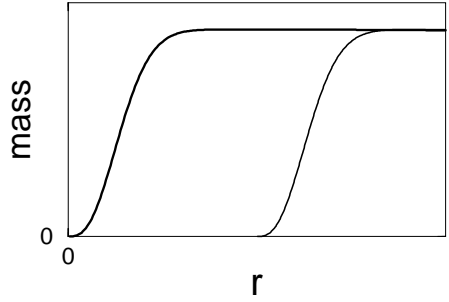}
\end{center}
\caption{The graphs of $m^*(r)$ (left) and $m(r)$.} \label{fig1}
\end{figure}

Using solutions (\ref{E:flat}) and (\ref{E:lambda}) in the
Einstein field equations, one can readily get the following
expressions for $p_r$ and $p_t$:
\begin{equation}\label{E:radial}
p_r = \frac{1}{8 \pi G  }\frac{l}{r^{2}}\left\{
  \left[1-\frac{4M}{\sqrt{\pi}r}
  ~\gamma \left( \frac{3}{2} ,
  \frac{(r-R_0)^2}{4\alpha}\right)\right]
   (1+l)-1\right\}
\end{equation}
(since $\nu^\prime = l/r$) and
\begin{equation}
p_t = \frac{1}{8 \pi G }\left[ \frac{1}{2}\left( 1 - \frac{2 m(r)
}{r} \right)\left( \frac{1}{2}(\nu^{\prime})^{2}+\nu^{\prime
\prime}-\frac{1}{2}\lambda^{\prime}\nu^{\prime}+\frac{1}{r}({\nu^{\prime
}-\lambda^{\prime}})\right)\right].
\end{equation}
Moreover, since $\lambda^\prime = \frac{2}{r} \frac{m^\prime r -
m}{r-2m} \quad$
  and $\quad m^\prime =
\frac{M(r-R_0)^2}{2\sqrt{\pi} ~\alpha^{\frac{3}{2}}}~exp
               \left(-\frac{(r-R_0)^2}{4\alpha}\right)$,
we arrive at
\begin{eqnarray}\label{E:transverse}
p_t = \frac{1}{8 \pi G}\frac{1}{4r^3}
  \left\{l^2 \left[r -
\frac{4M}{\sqrt{\pi}} ~\gamma \left( \frac{3}{2} ,
\frac{(r-R_0)^2}{4\alpha}\right)\right]\right.\nonumber \\ \left.
-2(l+2)\left[\frac{Mr(r-R_0)^2}
   {2\sqrt{\pi} \alpha^{3/2}}~exp
       \left(-\frac{(r-R_0)^2}{4\alpha}\right)
    - \frac{2M}{\sqrt{\pi}} \gamma \left( \frac{3}{2}.
\frac{(r-R_0)^2}{4\alpha}\right)\right]\right\}.
\end{eqnarray}

As a final comment, the spacetime is not asymptotically flat. If
this spacetime were to be joined to an exterior Schwarzschild
spacetime at the boundary of the halo, then the pressure
anisotropy would be an advantage. The reason is that, according to
Ref. \cite{kN09}, the solution cannot be matched to the
Schwarzschild exterior metric at the boundary if the pressure is
isotropic.

\section{Stability of circular orbits}\label{S:stability}
Given the four-velocity $U^{\alpha}=\frac{dx^{\sigma}}{d\tau}$
of a test particle moving solely in the subspace of the halo
and restricting ourselves to $\theta=\pi/2$, the equation
$g_{\nu\sigma}U^{\nu}U^{\sigma}=-m_{0}^{2}$ can be cast in a
Newtonian form
\begin{equation}
\left( \frac{dr}{d\tau}\right)  ^{2}=E^{2}+V(r),
\end{equation}
which gives

\begin{equation}
V(r)=-\left[E^{2}\left( 1-\frac{r^{-l}e^{-\lambda} }{B_{0}}\right)
+e^{-\lambda}\left( 1+\frac{L^{2}}{r^{2}}\right) \right].
 \end{equation}
The constants

\begin{equation}
E=\frac{U_{0}}{m_{0}}\quad and \quad L=\frac{U_{3}}{m_{0}}
\end{equation}
are, respectively, the conserved relativistic energy and angular
momentum per unit rest mass of the test particle \cite{kN09}.
Circular orbits are defined by $r=R_0$, a constant, so that
$\frac{dR_0}{d\tau}=0$ and, additionally,
$\frac{dV}{dr}\mid_{r=R_0}=0$. From these two conditions follow
the conserved parameters:
\begin{equation}
L=\pm\sqrt{\frac{l}{2-l}}R_0
\end{equation}
and, using $L$ in $V(R)=-E^{2}$, we get
\begin{equation}
E=\pm\sqrt{\frac{2B_{0}}{2-l}}R_0^{l/2}.
\end{equation}
The orbits will be stable if $\frac{d^{2}V}{dr^{2}}\mid_{r=R_0}<0$
and unstable if $\frac{d^{2}V}{dr^{2}}\mid_{r=R_0}>0$. We first
determine $\frac{d^{2}V}{dr^{2}}$ at $r=R_0$ by substituting the
expressions for $L$ and $E$:
\begin{eqnarray}\label{E:potential1}
   \left .\frac{d^{2}V}{dr^{2}}\right |_{r=R_0}=A+B+C+D,\nonumber
\end{eqnarray}

where
\begin{eqnarray}
A=\left[1-\frac{4M}{r\sqrt{\pi}}\int^{(r-R_0)^2/4\alpha}_0
  \sqrt{t}e^{-t}dt\right]\left[\frac{2}{2-l}l(l+1)\frac{R_0}{r^{2+l}}
      -\frac{6l}{2-l}\frac{R_0^2}{r^4}\right],\nonumber
\end{eqnarray}

\begin{eqnarray}
  B=-\frac{2M}{r^3}\left[\frac{r^2(r-R_0)}
    {\sqrt{\pi}\alpha^{3/2}}e^{-(r-R_0)^2/4\alpha}
   -\frac{r^2(r-R_0)^3}{4\sqrt{\pi}\alpha^{5/2}}
      e^{-(r-R_0)^2/4\alpha}\right. \nonumber \\
      \left.-\frac{r(r-R_0)^2}{\sqrt{\pi}\alpha^{3/2}}
      e^{-(r-R_0)^2/4\alpha}\right]
      \left(\frac{2}{2-l}\frac{R_0}{r^l}-1-\frac{l}{2-l}
     \frac{R_0^2}{r^2}\right),\nonumber
\end{eqnarray}

\begin{eqnarray}
    C=-\frac{4M}{r^2}\frac{r(r-R_0)^2}{2\sqrt{\pi}\alpha^{3/2}}
          e^{-(r-R_0)^2/4\alpha}
      \frac{2}{2-l}l\left(-\frac{R_0}{r^{1+1}}
     +\frac{R_0^2}{r^3}\right),\nonumber
\end{eqnarray}

\begin{eqnarray}
     D=\frac{8M}{\sqrt{\pi}r^2}\left[\int^{(r-R_0)^2/4\alpha}_0
    \sqrt{t}e^{-t}dt\right]\frac{2}{2-l}l
     \left(-\frac{R_0}{r^{1+l}}+\frac{R_0^2}{r^3}\right)\nonumber \\
     -\frac{8M}{\sqrt{\pi}r^3}\int^{(r-R_0)^2/4\alpha}_0
        \sqrt{t}e^{-t}dt
     \left(\frac{2}{2-l}\frac{R_0}{r^l}-1-\frac{l}{2-l}
    \frac{R_0^2}{r^2} \right).
\end{eqnarray}

In Eqs. (\ref{E:density1}) and (\ref{E:mass1}), $r$ is the
distance in the radial outward direction. We will therefore assume
that $r>R_0$, which does not result in a loss of generality. Our
aim is to show that under certain conditions, $V''(R_0)$ is
negative, resulting in a stable orbit.

Since $\alpha$ controls the ``width" of the Gaussian curve, we
study the smearing effect by assuming that
$r-R_0\approx\sqrt{\alpha}$, the fundamental condition discussed
in Sec. \ref{S:solutions}. Returning to Eq. (\ref{E:potential1}),
the mass of the shell, which we are treating as the analogue of a
smeared central object, has a relatively small value when measured
in kpc (and we must therefore assume that $G=1$). In conjunction
with the condition $r-R_0\approx\sqrt{\alpha}$, the first term $A$
is less than $l/R_0$. Next, factoring $B$, we have
\begin{eqnarray}
  \left[r-\frac{r(r-R_0)^2}{4\alpha}-(r-R_0)\right]
   \left(-\frac{2M}{r^2}\frac{r-R_0}
      {\sqrt{\pi}\alpha^{3/2}}e^{-(r-R_0)^2/4\alpha}
    \right) \nonumber \\
     \times \left(\frac{2}{2-l}\frac{R_0}{r^l}
    -1-\frac{l}{2-l}\frac{R_0^2}{r^2} \right).\nonumber
\end{eqnarray}
Since $(r-R_0)^2\approx\alpha$, the first factor is approximately
equal to $R_0-r/4$, which is positive for $r$ not too large.
Observe that $B$ exceeds $C$ in absolute value.  Because of the
$\alpha$ in the denominator, $|B|$ easily overtakes $A$ as well,
even for moderately small $\alpha$. Finally, since $D$ is
negative, $V''(R_0)$ is also negative, provided, of course, that
$R_0-r/4$ is positive.  The last condition limits the size of the
variable $r$.

\subsection{The effect of the noncommutative geometry}
It is shown by Nicolini, Smailagic and Spallucci~\cite{NSS06} that
at large distances one would expect only a minimal deviation from
the standard vacuum Schwarzschild geometry. Referring to Eqs.
(\ref{E:density1}) and (\ref{E:mass1}), new physics can only be
expected if $r\approx \sqrt{\alpha}$. In our model, this condition
corresponds to $r-R_0\approx\sqrt{\alpha}$. As we just saw, if $r$
is too large, then $V''(R_0)$ is no longer negative. This case is
an example of a large distance that produces the appearance of a
Schwarzschild geometry since the smearing, although very much
present, is no longer apparent. The reason is that, given the
present state of research in this field, it is not clear whether
the value of alpha can be determined. An interesting consequence
of this lack is that, in a sense, the unseen dark matter is
replaced by the unseen noncommutative geometry. (Regarding
large-scale structures, it is worth noting that a Gaussian source
has also been used in Ref. \cite{sS05} to model phantom-energy
supported wormholes, as well as in Ref. \cite{NS10} to model the
physical effects of short-distance fluctuations of noncommutative
coordinates in the study of black holes).

Let us now come to the issue of reproduction of flat rotation
curves from our model. It can be observed that the equation
(\ref{E:flat}) has also been employed by Farook et al. \cite{fR08}
(vide the equation (9) therein). Therefore, based on the
discussion made in the beginning of the Sec. 3 of the present
investigation and following Farook et al. \cite{fR08}, we can
reproduce the behaviours of the present model in the Fig. 2. It is
just a graph to illustrate that the  correct shape can be achieved
by using the following ansatz $v_{\phi}$ of the form
$v_{\phi}=\alpha
r\,\exp(-k_{1}r)+\beta\left[1-\exp(-k_{2}r)\right]$. Note that the
constants $\alpha$ and $\beta$ in this ansatz can have any units
we want, and hence the whole graph can be arbitrarily re-scaled. A
simpler way to put it is that the parameters $\alpha$ and $\beta$
in the ansatz can be chosen to have values that fit the
observational data. If one wants the graph to be quantitatively
correct (not just qualitatively) one just needs to let $\alpha
\rightarrow \alpha/3$ and  $\beta \rightarrow \beta/3$, but we did
not use the data for Fig. 2 in the rest of the paper. Just
re-scaling the parameters, one gets the desire figure. In a
similar way the behaviours, via Fig. 3, can be seen for $e^{\nu}$
vs $r$.

\begin{figure}
\begin{center}
\vspace{0.5cm}
\includegraphics[bb=10 86 502 654, scale=0.33, clip, height=5cm, width=5.3cm]{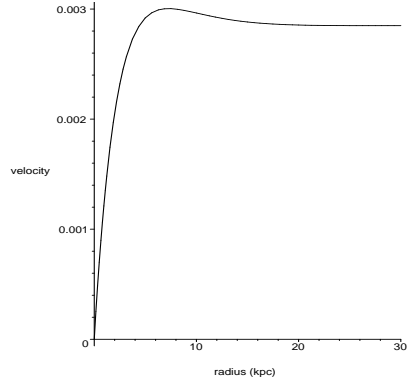}
\caption{A sample rotation curve utilizing an ansatz equation for
$v_{\phi}$ of the form $v_{\phi}=\alpha
r\,\exp(-k_{1}r)+\beta\left[1-\exp(-k_{2}r)\right]$.}
\label{fig:2}
\end{center}
\end{figure}

\begin{figure}
\begin{center}
\vspace{0.5cm}
\includegraphics[bb=21 26 440 497, scale=0.5, clip, keepaspectratio=true]{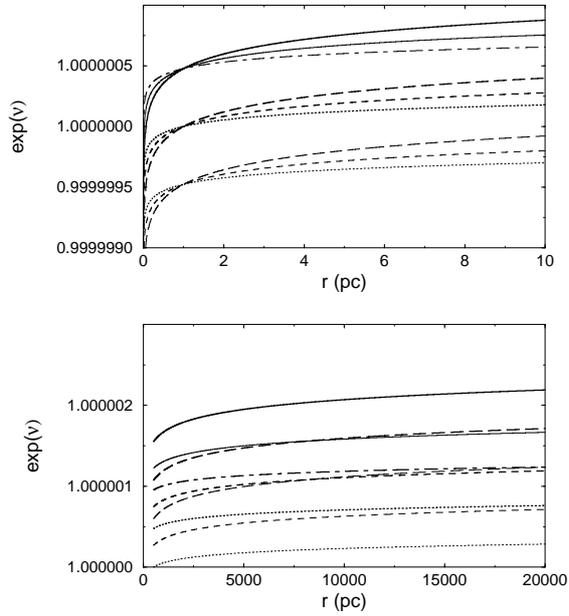}
\caption{Plot for the variation of $e^{\nu}$ vs $r$. The upper and
lower panels correspond to short and long $r$ behaviours. The
dotted, dashed and long-dashed curves represent $v_\phi=200$,
$250$ and $300$ Kms/second, respectively. For all these, thin
curves and thick curves represent $B_0=0.9999995$ and  $B_0=1.0$,
respectively. The chain, solid and thick solid curves
respectively, represent $v_\phi=200$, $250$ and $300$ Kms/second
but for $B_0=1.0000005$. } \label{fig:3}
\end{center}
\end{figure}

\subsection{Negative radial pressure}
It is emphasized in Ref. \cite{NSS06} that a negative radial
pressure is needed to retain the smearing effect near the origin,
referring now to Eqs. (\ref{E:density1}) and (\ref{E:mass1}). The
negative pressure counteracts the inward gravitational pull,
thereby preventing a collapse to a matter point.

Returning to Eq. (\ref{E:radial}), if $r-R_0\approx\sqrt{\alpha}$,
then $p_r$ is indeed negative if $M$, the mass of the shell, is
sufficiently large (since the shell need not be arbitrarily thin).
But as long as $r-R_0\approx\sqrt{\alpha}$, we have $V''(R_0)<0$,
so that the negative pressure is attributable to the
noncommutative geometry, much like traversable wormholes sustained
in this manner \cite{LG08}. We will return to this observation in
Sec. \ref{S:EoS}.

\section{Attraction and total gravitational energy}
Our next step is to consider the question of attractive gravity by
studying the geodesic equation for a test particle that is moving
along a circular path of radius $r=R$:
\begin{equation}\label{E:geodesic}
           \frac{d^2x^\alpha}{d\tau^2}
  +\Gamma_{\mu\gamma}^{\alpha}\frac{dx^\mu}
   {d\tau}\frac{dx^\gamma}{d\tau}= 0.
          \end{equation}
This equation implies that
\begin{equation}
  \frac{d^2 r} {d\tau^2} = - \frac{1}{2} e^{-\lambda}
    \left[\frac{d }{dr}e^\lambda \left(\frac{dr}{d\tau}
\right)^2 + \frac{d }{dr}e^\nu\left(\frac{dt}{d\tau}
\right)^2\right],
\end{equation}
making use of Eq. (\ref{E:flat}). As before, as long as
$\frac{dr}{d\tau}=0$, we get
\begin{equation}
  \frac{d^2 r} {d\tau^2} = - \frac{1}{2} e^{-\lambda}
   B_0lr^{l-1}\left(\frac{dt}{d\tau}\right)^2<0.
\end{equation}
We conclude that objects are attracted toward the center.

According to Lyndell-Bell et al. \cite{dL07}, we can determine the
total gravitational energy $E_G$ between two fixed radii $r_1$ and
$r_2$ by means of the following formula:
\begin{equation} E_{G}=M_N-E_{M}=4\pi\int_{r_{1}}^{r_{2}}
    [1-\sqrt{e^{\lambda(r)}}]\rho r^{2}dr,
\end{equation}
where
\begin{equation}
M_N=4\pi\int_{r_{1}}^{r_{2}}\rho r^{2}dr
\end{equation}
is the Newtonian mass given by

\begin{equation}
M_N = 4 \pi \int_{r_{1}}^{r_{2}} \rho r^2 dr =
   \frac{M}{(4 \pi)^{1/2}
  (\alpha)^{\frac{3}{2}}}\int_{r_{1}}^{r_{2}}
    exp\left( - \frac{
    (r-R_0)^2}{4 \alpha} \right) r^2 dr.
\end{equation}
So the total gravitational energy is
\begin{equation} E_{G}=\frac{M}{(4 \pi)^{1/2}
(\alpha)^{\frac{3}{2}}}\int_{r_{1}}^{r_{2}} \left[1 - \left[1 -
\frac{2 m(r)}{r}\right]^{- \frac{1}{2}}\right] exp \left( -
\frac{(r-R_0)^2}{4 \alpha} \right) r^2 dr,
\end{equation}
where $m(r)$ is given in Eq. (\ref{E:mass2}). As noted in Sec. 4,
$M$ is relatively small. So the integrand, and hence $E_G$, are
negative, showing that gravity in the halo is indeed attractive.

\section{The observed equation of state}\label{S:EoS}
The equation of state of the halo fluid can be obtained from a
combination of rotation curves and lensing measurements. To this
end, let us rewrite the metric, Eq. (\ref{E:line1}), in the
following form:
\begin{equation}
ds^2 = - e^{2\Phi(r)} dt^2 + \frac{1}{[1 - \frac{2
m(r)}{r}]}dr^2+r^2+r^{2}(d\theta^{2} + sin^{2}\theta d\phi^{2}),
\end{equation}
where
\begin{equation}
\Phi(r) =  \frac{1}{2} \left[ \ln B_0 + l \ln r \right]
\end{equation}
and $m(r)$ is given in Eq. (\ref{E:mass2}). As discussed in Ref.
\cite{kN09}, the functions are determined indirectly from certain
lensing measurements defined by
\begin{equation}
\Phi_{lens}=\frac{\Phi(r)}{2}+\frac{1}{2}\int\frac{m(r)}{r^{2}}dr
= \frac{\ln B_0}{4} +\frac{ \ln r^l}{4} + \int\frac{M~ \gamma
\left(3/2,(r-R_0)^2/4\alpha\right)}{\sqrt{\pi}r^2}dr
\end{equation}
and
\begin{equation}
m_{lens}=\frac{1}{2}r^{2}\Phi^{\prime}(r)+\frac{1}{2}m(r)
=\frac{lr}{4} + \frac{M ~\gamma \left( 3/2 ,
(r-R_0)^2/4\alpha\right)}{\sqrt{\pi}}.
\end{equation}
The observed equation of state depends on the dimensionless quantity
\begin{equation}
\omega(r)=\frac{p_r+2p_t}{3\rho}\approx
\frac{2}{3}\frac{m_{RC}^{\prime}-m_{lens}^{\prime}}{2m_{lens}^{\prime}-m_{RC}^{\prime}},
\end{equation}
due to Faber and Visser \cite{FV06}. The subscript $RC$ refers to
the rotation curve, i. e.
\begin{equation}
\phi_{RC} = \Phi(r)= \frac{1}{2} \left[ \ln B_0 + l \ln r \right]
\end{equation}
and
\begin{equation}
m_{RC} = r^2\Phi^{\prime}(r)= \frac{l r}{2}.
\end{equation}
The prime denotes the derivative with respect to $r$. The result
is
\begin{equation}
\omega(r)=\frac{2}{3}\frac{m_{RC}^{\prime}-m_{lens}^{\prime}}{2m_{lens}^{\prime}-m_{RC}^{\prime}}
=\frac{l\sqrt{\pi}\alpha^{\frac{3}{2}} - M (r-R_0)^2 exp\left[ -
(r-R_0)^2/4 \alpha \right]} {3 M (r-R_0)^2 ~exp\left[-(r-R_0)^2/4
\alpha \right]}.
\end{equation}

At first glance, $\omega(r)>0$ due to the small value of $M$, just
as in the case of ordinary matter. But if $\alpha$ is very small
and $M$ sufficiently large, a situation we encountered in Sec.
\ref{S:stability}, then $\omega(r)<0$.  This also follows from the
fact that $p_t<0$ for a sufficiently small $\alpha$ [Eq.
(\ref{E:transverse})], since we already know that $p_r$ is
negative. So the ``quintessence-like" condition is due entirely to
the noncommutative geometry.  Observe, however, that as $\alpha$
gets very small, $\omega(r)$ approached $-1/3$ from the right, so
that $\omega(r)$ is not in the actual quintessence range.

As a final comment, at large distances, where the pressures are no
longer negative, we have $\omega(r)>0$. This may be interpreted to
mean that the smearing, although still present, is no longer seen
at this distance, resulting in the appearance of ordinary matter.

\section{Conclusion}
In this paper we assume the existence of flat rotation curves in
galaxies and discuss such curves in the context of noncommutative
geometry. The energy density of this geometry is a smeared
gravitational source, where the mass is diffused throughout a
region of linear dimension $\sqrt{\alpha}$. This diffusion is due
to the uncertainty attributable to the commutator and is
applicable to other large-scale structures. It is shown that this
intrinsic property of the noncommutative geometry is able to
account for the stable circular orbits in remote regions of the
halo, as well as for attractive gravity. Sufficiently far from the
orbit, the smearing is no longer observed, which may be
interpreted to mean that stable orbits are due to the unseen
noncommutative geometry instead of the unseen dark matter.

\section*{Acknowledgement}
The authors are indebted to Vance Gladney and Andrew
DeBenedictisfor for many helpful and illuminating discussions. FR,
AAU and SR are specially thankful to the authority of
Inter-University Centre for Astronomy and Astrophysics, Pune,
India for providing them Visiting Associateship under which a part
of this work was carried out. FR is also thankful to PURSE for
providing financial support. Thanks are also due to anonymous
referee for imparting valuable suggestions which have enabled us
to improve the manuscript.

\end{document}